\documentclass[reprint,superscriptaddress,preprintnumbers,amsmath,amssymb,aps,floatfix,longbibliography,amsfonts]{revtex4-2}

\usepackage[utf8]{inputenc}
\usepackage{graphicx}
\usepackage{dcolumn}
\usepackage{bm}
\usepackage[colorlinks=true
,urlcolor=blue
,anchorcolor=blue
,citecolor=blue
,filecolor=blue
,linkcolor=red
,menucolor=blue
,linktocpage=true
,pdfproducer=medialab
,pdfa=true
]{hyperref}
\usepackage[mathlines]{lineno}
\usepackage{xspace}
\usepackage{bbding}
\usepackage{siunitx}
\usepackage{tikz,xcolor}
\usepackage{comment}
\usepackage{amsmath}

\usepackage{physics2}
\usephysicsmodule{ab}   
\usephysicsmodule{ab.braket}    
\usephysicsmodule{diagmat}  
\usephysicsmodule[showleft=2,showtop=2]{xmat}   

\newcommand{\thistitle}{Search for Dark Photon Dark Matter with a Mass around $\SI{36.1}{\micro eV}$ Using a Frequency-tunable Cavity Controlled through a Coupled Superconducting Qubit \xspace}

\begin{document}

\title{\thistitle}

\author{Kan Nakazono}
\email[Correspondence to: ]{nakazono@icepp.s.u-tokyo.ac.jp}
\affiliation{Department of Physics, Graduate School of Science, The University of Tokyo, Tokyo 113-0033, Japan}
\affiliation{International Center for Quantum-field Measurement Systems for Studies of the Universeand Particles (QUP, WPI), High Energy Accelerator Research Organization (KEK), Oho 1-1, Tsukuba, Ibaraki 305-0801, Japan}
\author{Shion Chen}
\affiliation{Department of Physics, Graduate School of Science, Kyoto University, Kyoto 606-8502, Japan}
\author{Hajime Fukuda}
\affiliation{Department of Physics, Graduate School of Science, The University of Tokyo, Tokyo 113-0033, Japan}
\author{Yutaro Iiyama}
\affiliation{International Center for Elementary Particle Physics (ICEPP), The University of Tokyo, Tokyo 113-0033, Japan}
\author{Toshiaki Inada}
\affiliation{International Center for Elementary Particle Physics (ICEPP), The University of Tokyo, Tokyo 113-0033, Japan}
\author{Takeo Moroi}
\affiliation{Department of Physics, Graduate School of Science, The University of Tokyo, Tokyo 113-0033, Japan}
\affiliation{International Center for Quantum-field Measurement Systems for Studies of the Universeand Particles (QUP, WPI), High Energy Accelerator Research Organization (KEK), Oho 1-1, Tsukuba, Ibaraki 305-0801, Japan}
\author{Tatsumi Nitta}
\affiliation{International Center for Quantum-field Measurement Systems for Studies of the Universeand Particles (QUP, WPI), High Energy Accelerator Research Organization (KEK), Oho 1-1, Tsukuba, Ibaraki 305-0801, Japan}
\author{Atsushi Noguchi}
\affiliation{RIKEN Center for Quantum Computing (RQC), 2-1 Hirosawa, Tokyo 351-0198, Japan}
\affiliation{Komaba Institute for Science (KIS), The University of Tokyo, Tokyo 153-8902, Japan}
\affiliation{Inamori Research Institute for Science (InaRIS), Kyoto 600-8411, Japan}
\author{Ryu Sawada}
\affiliation{International Center for Elementary Particle Physics (ICEPP), The University of Tokyo, Tokyo 113-0033, Japan}
\author{Shotaro Shirai}
\affiliation{RIKEN Center for Quantum Computing (RQC), 2-1 Hirosawa, Tokyo 351-0198, Japan}
\affiliation{Komaba Institute for Science (KIS), The University of Tokyo, Tokyo 153-8902, Japan}
\author{Thanaporn Sichanugrist}
\affiliation{Department of Physics, Graduate School of Science, The University of Tokyo, Tokyo 113-0033, Japan}
\author{Koji Terashi}
\affiliation{International Center for Elementary Particle Physics (ICEPP), The University of Tokyo, Tokyo 113-0033, Japan}
\author{Karin Watanabe}
\affiliation{Department of Physics, Graduate School of Science, The University of Tokyo, Tokyo 113-0033, Japan}
\affiliation{International Center for Quantum-field Measurement Systems for Studies of the Universeand Particles (QUP, WPI), High Energy Accelerator Research Organization (KEK), Oho 1-1, Tsukuba, Ibaraki 305-0801, Japan}

\collaboration{DarQ Collaboration}\noaffiliation


\begin{abstract}
We report the results of a search for dark photon dark matter using a cavity that employs a transmon qubit as a frequency tuner. The tuning mechanism utilizes the energy level shift arising from the mode mixing between the qubit and the cavity mode. This method is advantageous as it avoids the frictional heating and electromagnetic leakage associated with mechanical tuning. We searched for a dark matter signal in the mass range 36.132 -- 36.179~$\si{\micro eV}$ and found no significant evidence. As a result, we set the exclusion limit on the kinetic mixing parameter down to approximately $5 \times 10^{-13}$, surpassing the existing limit set by cosmology.
\end{abstract}

\maketitle
\textit{Introduction.---}
Dark matter (DM) is a central subject in both cosmology and astrophysics, yet its particle nature has not been confirmed experimentally. One prominent class of DM candidates is wave-like DM, whose mass lies in the sub-eV range or below \cite{butler2023report2021uscommunity,huiWaveDarkMatter2021}. Among these, the dark photon, a hypothetical massive vector field that kinetically mixes with the ordinary photon, is a particularly well-motivated and widely studied candidate. 

Various experimental approaches have been developed to detect the dark photon DM \cite{sikivieExperimentalTestsInvisible1983,hornsSearchingWISPyCold2013, ehretNewALPSResults2010,PhysRevLett.128.131801}, and among them, cavity haloscope experiments \cite{sikivieExperimentalTestsInvisible1983} currently offer the highest sensitivity. In those experiments, sensitivity is enhanced when the cavity resonance frequency matches the frequency of photons converted from DM; this frequency corresponds to the DM mass. Since the DM mass is unknown, the cavity frequency must be scanned to probe a wide mass range. The signal bandwidth, determined by the DM velocity dispersion, is typically six orders of magnitude smaller than the considered viable DM mass range, making frequency scan one of the biggest challenges in the resonance-based haloscope searches. The most commonly adopted method for tuning the frequency is by physically varying the boundary conditions of the cavity using metallic rods \cite{khatiwadaAxionDarkMatter2021,PhysRevLett.126.191802}, dielectric materials \cite{PhysRevD.106.052007}, or other movable elements \cite{PhysRevD.107.055013}. These methods require complex mechanical structures, such as stepper motors with cryogenic gears or piezoelectric actuators, that operate at low temperatures. Such components introduce considerable engineering complexity and can generate frictional heat, which degrades sensitivity and limits the scan speed. In addition, electromagnetic leakage in the cavity can compromise the quality factor of the resonance.

To address these challenges with frequency tuning, we present the first results from the DarQ-Lamb experiment, conducted by the DarQ collaboration \footnote{\url{https://sites.google.com/view/darq-experiment/home}}. This work reports the first search for dark photon dark matter in the previously unexplored mass window around \SI{36.1}{\micro eV}, using a cavity whose frequency is tuned via a coupled qubit. Our work is part of a growing trend of integrating quantum technologies into wave-like DM searches, including cavity haloscope experiments. This has included enhancing readout sensitivity with devices like Josephson Parametric Amplifiers (JPAs) \cite{yamamotoFluxdrivenJosephsonParametric2008,PhysRevLett.124.101303} and Traveling Wave Parametric Amplifiers (TWPAs) \cite{macklinQuantumLimitedJosephson2015,divoraSearchGalacticAxions2023}, as well as exploring superconducting qubits as quantum sensors capable of surpassing the standard quantum limit \cite{PhysRevLett.126.141302,PhysRevLett.131.211001,PhysRevD.110.115021,PhysRevX.15.021031}. This study uses the qubit technology in a different direction, employing it as a frequency tuning system. A search using a similar tuning mechanism was reported in Ref.~\cite{zhaoFluxTunableCavityDark2025} in a different mass range. Our implementation, however, features notable differences. While the experiment in Ref.~\cite{zhaoFluxTunableCavityDark2025} utilized a cavity with an internal line for flux biasing and employed a separate qubit as a quantum detector, our setup takes a simple approach, using an external coil for flux control and a standard amplifier chain for signal readout.

\textit{Theoretical Background.---} 
The frequency tuning mechanism relies on the mixing of the qubit and the cavity mode described by the Jaynes-Cummings Hamiltonian \cite{jaynesComparisonQuantumSemiclassical1963}:
\begin{align}
  \hat{H}_{\rm JC}^{(0)} = \hbar \omega_c \hat{a}^\dagger \hat{a}
  + \frac{\hbar}{2} \omega^{(0)}_q \hat{\sigma}_z
  - g \hbar ( \hat{a} \hat{\sigma}_+ + \hat{a}^\dagger \hat{\sigma}_- ),
\end{align}
where $\hat{a}$ and $\hat{a}^\dagger$ are the annihilation and creation operators of the cavity photon mode of interest. $g$ is the coupling constant between the cavity and the qubit. The Pauli operators for the qubit are denoted as $\hat{\sigma}_{x,y,z}$, and we define $\hat{\sigma}_\pm \equiv \frac{1}{2}(\hat{\sigma}_{x} \pm i\hat{\sigma}_{y})$.

We apply the following unitary transformation: $\hat{H}_{\rm JC} \equiv \hat{U} \hat{H}_{\rm JC}^{(0)} \hat{U}^\dagger$, where the unitary operator $\hat{U}$ is given by $\hat{U} \equiv \exp[ -\frac{g}{ \Delta} ( \hat{a} \hat{\sigma}_+ - \hat{a}^\dagger \hat{\sigma}_- ) ]$, with $\Delta \equiv \omega_q - \omega_c$ being the detuning. This transformation yields

\begin{align} \label{eq:JCeq1}
\hat{H}_{\rm JC}&=\hbar\ab(\omega_c+\frac{g^2}{\Delta} \hat{\sigma}_z)\hat{a}^{\dagger}\hat{a}+\frac{\hbar}{2}\ab(\omega_q^{(0)}+\frac{g^2}{\Delta})\hat{\sigma}_z + {O}(g^3).
\end{align}
This result shows that the cavity resonance frequency is shifted by $\pm \frac{g^2}{\Delta}$ depending on the qubit state. Importantly, since the detuning $\Delta$ changes with the qubit frequency $\omega_q$, the cavity resonance frequency becomes tunable via control of $\omega_q$.

 Dark photons mix kinetically with ordinary photons via the kinetic mixing described by the following interaction Lagrangian:
\begin{align}
  \mathcal{L}_{\rm int} = \frac{1}{2} \chi F_{\mu\nu} X^{\mu\nu},
\end{align}
where $\chi$ is the kinetic-mixing parameter, and $F_{\mu\nu}$ ($X_{\mu\nu}$) is the field strength tensor of the ordinary (dark) photon. Such an interaction enables the conversion of dark photons into ordinary photons within a resonant cavity. When the dark photon mass matches the cavity resonance frequency, the dark photon field resonantly excites the cavity mode, leading to an accumulation of electromagnetic energy in the cavity. This energy can be extracted and measured through an antenna. The expected signal power from dark photon DM is given by \cite{ariasWISPyColdDark2012} :

\begin{align}
  P_{\rm signal} = \eta \chi^2 m_{\gamma'} \rho_{\rm DM} V_{\rm eff} Q_{\rm L} 
  \frac{\beta}{1+\beta} L(f, f_0, Q_{\rm L}), \label{eq:Ps}
\end{align}
where $\eta$ is the attenuation factor that accounts for signal loss from the subtraction of gradual background fluctuations during the digital filtering process \cite{cervantesADMXOrpheusFirstSearch2022}, $m_{\gamma'}$ is the dark photon mass, $\rho_{\rm DM}$ is the local DM energy density (assuming DM consists entirely of dark photons), $\beta$ is the antenna coupling coefficient between the cavity and the antenna, $Q_{\rm L}$ is the loaded quality factor of the cavity, and $L(f, f_0, Q_{\rm L})$ is a Lorentzian function representing the resonance response of the cavity, defined as
\begin{align}
  L(f, f_0, Q_{\rm L}) = \frac{1}{1 + 4 Q_{\rm L}^2 \left( \frac{f - f_0}{f_0} \right)^2 }.
\end{align}
Here, $f$ is the dark photon signal frequency and $f_0$ is the cavity resonance frequency. The effective volume $V_{\rm eff}$, which characterizes the overlap between the dark photon polarization and the cavity field, is given by
\begin{align}
  V_{\rm eff} \equiv
  \frac{\left| \int dV\, \vec{E}_c \cdot \vec{n} \right|^2}
       {\int dV\, |\vec{E}_c|^2},
\end{align}
where $\vec{E}_c$ is the electric field profile of the cavity mode of interest, $\vec{n}$ is the polarization vector of the dark photon DM, and the integration is performed within the cavity volume.
The detection sensitivity is determined by the signal-to-noise ratio (SNR). Using the Dicke radiometer equation \cite{Dicke1946}, the SNR for a specific frequency bin is defined as
\begin{align} \label{eq:SNR}
    \mathrm{SNR_{\rm bin}} \equiv \frac{P_{\rm signal, bin}}{P_{\rm noise, bin}}\sqrt{N_{\rm bin}},
\end{align}
where $P_{\rm signal, bin}$ is the signal power in the bin, $P_{\rm noise, bin}$ is the noise power in the bin, and $N_{\rm bin}$ is the total number of measurements performed for that bin.
The dominant source of noise in our setup is expected to be Johnson–Nyquist noise  \cite{johnsonThermalAgitationElectricity1928,PhysRev.32.110} of the first amplifier. The noise power is characterized as
\begin{align}\label{eq:pnoise}
  P_{\rm noise} = b k_{\rm B} T_{\rm sys},
\end{align}
where $b$ is the bandwidth of the one bin, $k_{\rm B}$ is the Boltzmann constant, and $T_{\rm sys}$ is the system noise temperature. 
In analysis section, this SNR is used as a criterion for detecting the dark photon signal.

\begin{figure}[tb]
  \centering
    \includegraphics[keepaspectratio,width=1.0\linewidth]{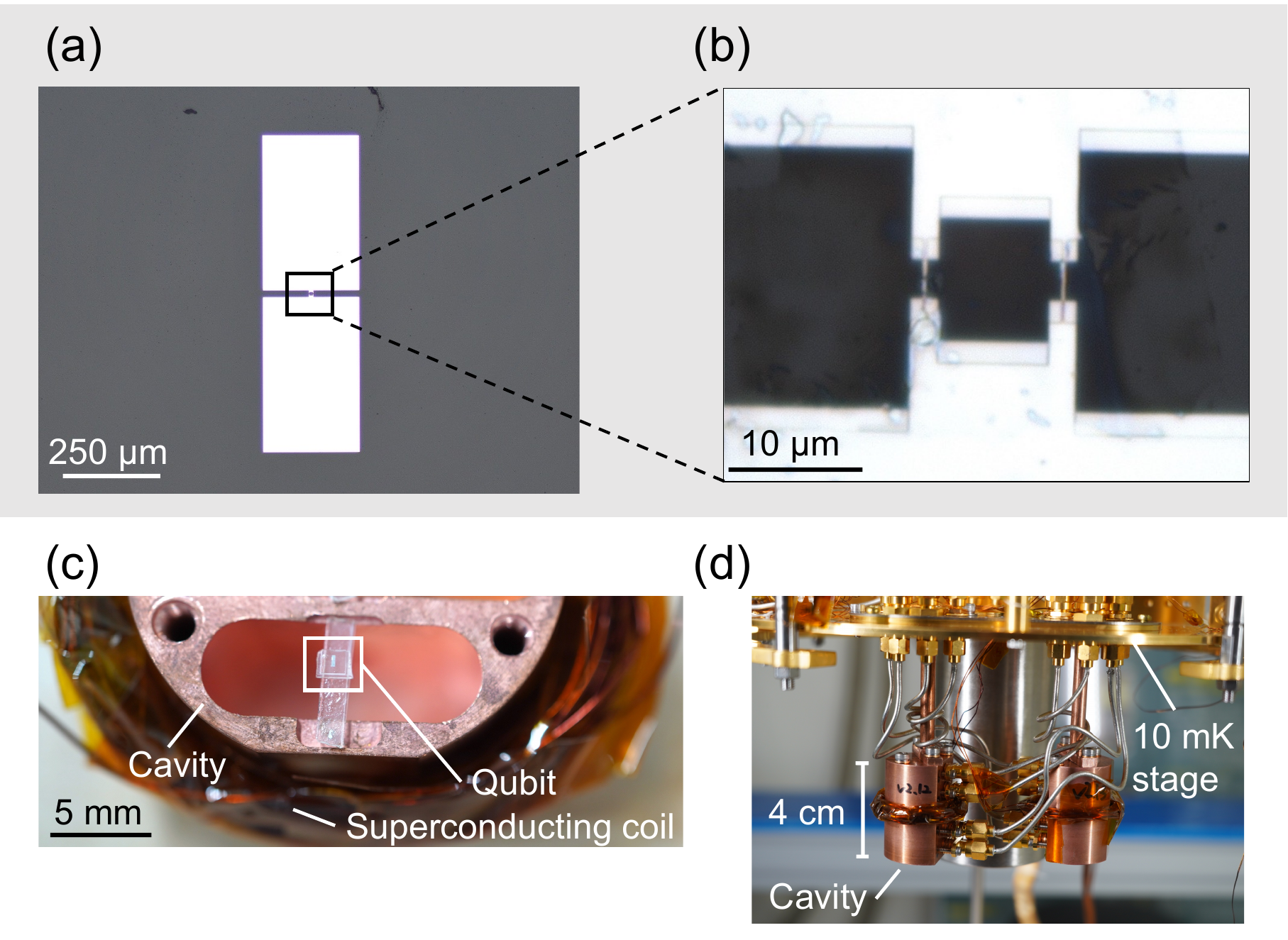}
    \caption{The experimental setup for the superconducting qubit and cavity. (a) Optical microscope image of the qubit used in this experiment. The scale of the large rectangle capacitance pads is $\SI{250}{\micro m}\times\SI{400}{\micro m}$. (b) The magnification of (a) capturing the SQUID loop in the qubit. (c) Photo of the copper cavity and the qubit in this experiment. The dimension of the cavity is $\SI{7}{mm} \times \SI{13}{mm} \times \SI{38}{mm}$. The qubit is on a bridge of sapphire and in the center of the cavity. The cavity is surrounded by a NbTi superconducting coil to impose magnetic flux. (d) Photo of the cavity being loaded onto the \SI{10}{mK} stage of the dilution refrigerator. The refrigerator is located at the cryogenic research center, the University of Tokyo. To protect against the ambient magnetic flux, the sample is encapsulated by a mu-metal based magnetic shield.}
    \label{fig:qubit}
\end{figure}

\textit{Methodology.---}
Fig.~\ref{fig:qubit} shows a superconducting qubit, a resonant cavity, and other components in our setup. This qubit (Fig.~\ref{fig:qubit}(a) and (b)) was fabricated with aluminum thin film on a sapphire substrate \footnote{C-plane [0001], purity 99.99\%, diameter:\SI{100}{mm}, and thickness:\SI{650}{um}}. The Josephson junction was fabricated using the Niemeyer-Dolan bridge technique \cite{dolanOffsetMasksLiftoff1977,10.1063/1.89094}. The process was done  in the Takeda super cleanroom at the University of Tokyo, except for the double angle evaporation process by Plassys MEB 550S2-HV in the Class 1000 cleanroom at Okinawa Institute of Science and Technology. 

The cavity in this experiment was fabricated from oxygen free copper that allows external magnetic flux. The cavity has a dimension of \SI{7}{mm} $\times$ \SI{13}{mm} $\times$ \SI{38}{mm}  with rounded corners at both ends (Fig.~\ref{fig:qubit}(c)), with two ports for transmission and reflection measurements. Simulations were performed in ANSYS HFSS and the effective volume $V_{\rm eff}$ was calculated to be $\SI{3.14 \pm 0.02}{cm^3}$.

The qubit chip is placed on a sapphire bridge located at the center of the cavity to maximize the coupling with the electric field of the fundamental mode of the cavity, TE101. The cavity is surrounded by a superconducting coil for magnetic flux control (Fig.~\ref{fig:qubit}(c)). At room temperature, the resonant frequency of the cavity with the qubit installed was approximately \SI{8.706}{GHz} and the quality factor is approximately $Q_{\rm RT}\sim4400$. The cavity was then thermally attached to the \SI{10}{mK} stage of the dilution refrigerator (Fig.~\ref{fig:qubit}(d)). The qubit-cavity coupling constant, $g$, was estimated to be approximately $\SI{54.8}{MHz}$ from a resonator power scan measurement \cite{PRXQuantum.2.040202}.

\begin{figure}[tb]
  \centering
    \includegraphics[width=0.96\linewidth]{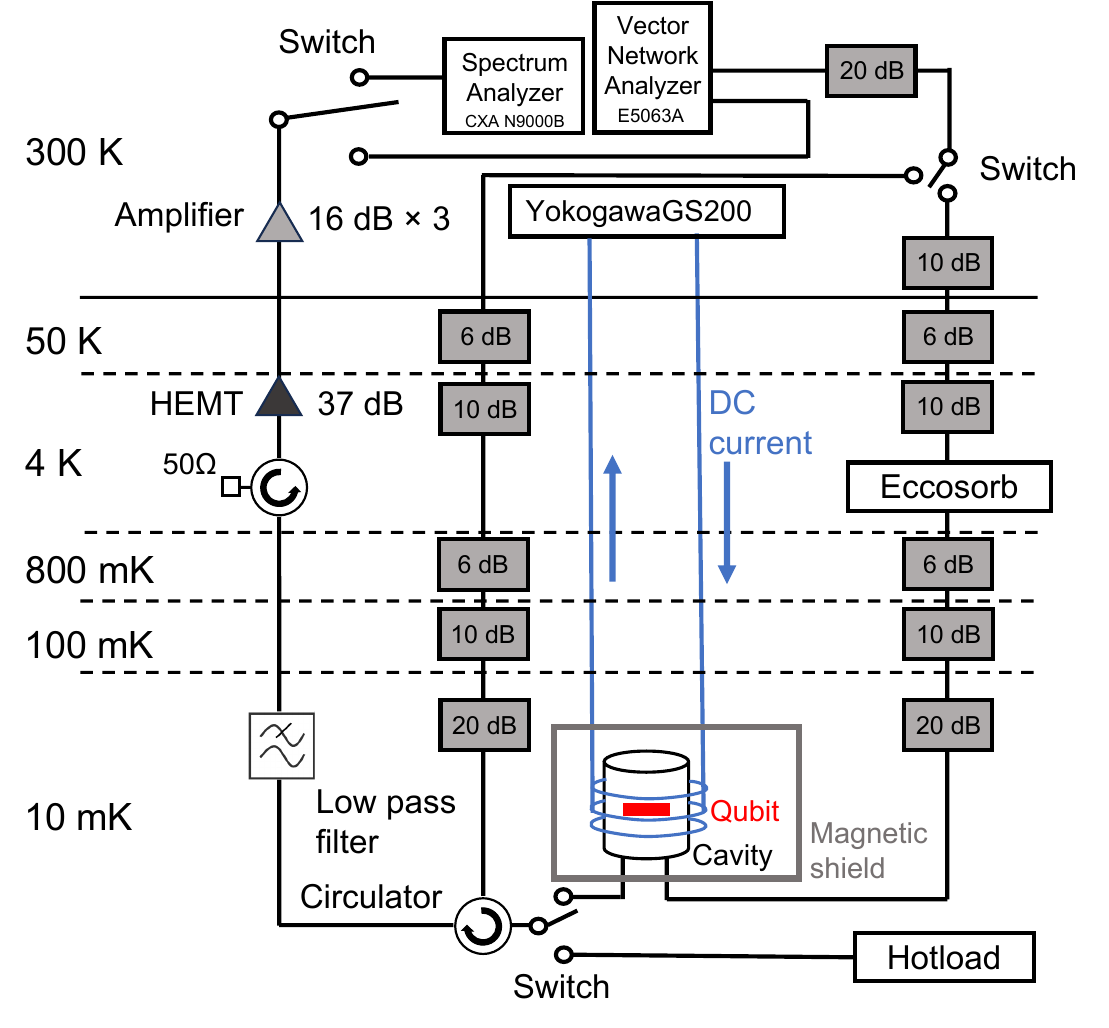}
    \caption{Schematic of the receiver chain used in this experiment. For the measurement, a spectrum analyzer (Keysight CXA N9000B) and a vector network analyzer (Keysight E5063A) were used. For the DC current modulation, Yokogawa GS200 was used as DC supply. The spectrum was amplified first by a HEMT and subsequently a series of room temperature amplifiers to detect weak signal. The switches at room temperature were used to switch between the transmission and reflection measurement and spectrum analyzer measurement and VNA measurement respectively. For the system noise calibration, a hotload was placed at the \SI{10}{mK} stage connected to the same RF circuit as the measured cavity.}
    \label{fig:reciver}
\end{figure}

The receiver chain in this experiment is shown in Fig.~\ref{fig:reciver}. The upper part of Fig.~\ref{fig:reciver}, at a temperature of \SI{300}{K}, represents the area outside the dilution refrigerator. This part contains a Vector Network Analyzer (VNA) for the transmission and the reflection measurements, a Spectrum Analyzer (SA) to measure the power spectrum that includes the signal from dark photons, a DC power supply to provide current for adjusting the magnetic flux through the qubit, a room-temperature amplifier, and switches to select ports for these instruments.

The lower part of the figure shows the interior of the dilution refrigerator. On the input RF line (the black line on the right and center), there are attenuators and an Eccosorb filter. The output line (the black line on the left side) is equipped with a circulator and a High-Electron-Mobility Transistor (HEMT), which serves as an amplifier. At the lowest temperature stage of \SI{10}{mK}, the cavity and the qubit are located within an electromagnetic shield.

In the dilution refrigerator, the temperature indicated by the thermometer on the cavity was around \SIrange[range-units = single,range-phrase = --]{50}{70}{mK} during the measurement.
In order to obtain the system noise temperature $T_{\rm sys}$, the $Y$-factor method \cite{wilson2011techniquesradioastronomy} was used. A \SI{50}{\ohm} terminator (``hotload'') was located before the directional coupler and the switch, to follow the same output path as the cavity (Fig.~\ref{fig:reciver}). The output power is measured as function of hotload temperature ($T_{\rm hotload}$), and $T_{\rm sys}$ was estimated by extrapolating the relation to $T_{\rm sys}=0$ assuming the linear relationship between the measured power and $T_{\rm hotload}$. As a result, $T_{\rm sys}$ was estimated as $\SI{7.9 \pm 0.6}{K}$. 

\begin{figure}[tb]
  \centering
    \includegraphics[width=0.96\linewidth]{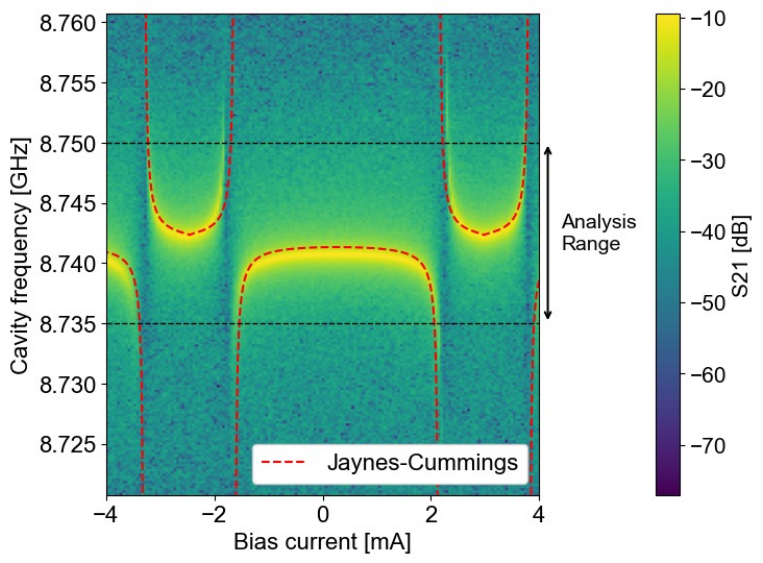}
    \caption{Relationship between the bias current flowing through the coil ($x$-axis) and the cavity frequency ($y$-axis). The color scale indicates the cavity transmittance, with the yellow region corresponding to the center of the cavity resonance as the function of the bias current. The red dashed line is the prediction by the Jaynes-Cummings model. The region marked by the black dashed lines and arrow corresponds to the approximate data range used in this study.}
    \label{fig:JC}
\end{figure}

 The measured cavity transmission property as a function of coil current is shown in Fig.~\ref{fig:JC}. This figure shows that the cavity frequency shifts when the bias current through the coil (Fig.~\ref{fig:qubit}(c)) varies. This behavior is well described by the Jaynes-Cummings model. The red dashed line in this figure is a fitted line of Jaynes-Cummings model, shown in Eq.~\ref{eq:JCeq1}.

The DM search was performed on Oct 2--4 2024, acquiring the digitized power spectra repeated 1540 times following the protocol.
\begin{enumerate}
  \item A transmission measurement was done by the VNA to determine the loaded quality factor $Q_{\rm L}$ and the resonant frequency of the cavity $f_{0}$.
  \item A reflection measurement was done by the VNA to determine the antenna coupling coefficient $\beta$.
  \item A power spectrum was measured by averaging 100 traces ($N_{\rm each}=100$) on a SA without any applied drive tone. The frequency span for the measurement was three times the cavity's full width at half maximum (FWHM$\sim\SI{1}{MHz}$).
  \item The DC current through the coil was adjusted to achieve a frequency scan step of about \SI{100}{kHz}. This step size was considerably smaller than the full width at half maximum.
\end{enumerate}

The measurement at a single frequency step took about 2--3 minutes. The relevant parameters on the measurement are summarized in Table~\ref{table:parameter}.

\begin{table*}[tb]
  \begin{center}
      \caption{Summary of relevant parameters to the search}
      \begin{tabular}{ccccc} \hline \label{table:parameter}
          Symbol &Description&Value &Uncertainty&Reference\\ \hline
          $f_{0}$ & Cavity frequency &$8.735 $--$\SI{8.750}{GHz}$&N/A& Cavity transmission measurement in each scan step \\
          $\rho_{\rm DM}$&Dark matter energy density&$\SI{0.45}{GeV/cm^3}$&N/A& Ref. \cite{asztalosLargescaleMicrowaveCavity2001}\\
          $\beta$ & Antenna coupling coefficient &$0.1$--$0.3$ &$2\%$& Cavity reflection measurement in each scan step\\
          $T_{\rm sys} $& System noise temperature&$\SI{7.9}{K}$& $7.6\%$&The $Y$-factor measurement\\
          $V_{\rm eff} $& Effective volume& $\SI{3.14}{cm^3}$ & $0.6\% $& ANSYS HFSS simulation\\
          $Q_{\rm L}$ & Loaded quality factor& $5000$--$10000$ &$4\% $& Cavity transmission measurement in each scan step\\
          $\eta$& Software attenuation factor & 1.02 &$7.7\%$& Extracted from Savitzky-Golay smooth curves\\ 
          $N_{\rm each}$& Number of measurement times & 100 &N/A& Number of averaging in each scan step\\
          $b$& Analyzer resolution bandwidth & $\SI{200}{Hz} $&N/A& Resolution bandwidth of spectrum analyzer\\   \hline
      \end{tabular}
  \end{center}
\end{table*}

\textit{Analysis.---}
\label{sec:analysis}We analyzed the acquired spectrum following a procedure established in previous cavity haloscope studies \cite{PhysRevD.96.123008,cervantesADMXOrpheusFirstSearch2022}.

First, as a preliminary data selection, measurements were discarded if the fit quality was poor, specifically when the reduced chi-squares satisfied $\chi_{\mathrm{trans}}^2 > 2$ or $\chi_{\mathrm{refl}}^2 > 5$. Data were also excluded if the extracted parameters fell outside the valid ranges: $\beta > 10$, $Q_{\mathrm{L}} < 1000$, or $Q_{\mathrm{L}} > 100000$. Furthermore, datasets with excessive relative errors in $\beta$ and $Q_{\mathrm{L}}$ were also rejected. Next, a Savitzky-Golay filter \cite{savitzkySmoothingDifferentiationData1964} was applied to mitigate baseline distortion originating from the receiver chain. This filtering was performed using a second-order polynomial with a window size of 1001.
The attenuation factor induced by this filter was determined to be $\eta=1.02\pm0.08$ by injection of software signal. Following the data selection and Savitzky-Golay filtering, the filtered power spectrum $P_{\mathrm{filtered}}$ was first normalized to have zero mean and unit standard deviation ($\mu=0$, $\sigma=1$) to obtain $P_{\mathrm{normalized}}$. This normalized spectrum was then rescaled by the noise power $P_{\mathrm{noise}}$ from Eq.~\ref{eq:pnoise} and the expected signal power $P(\chi=1)$ from Eq.~\ref{eq:Ps} as follows:
\begin{align}
    P_{\mathrm{rescaled}} \equiv P_{\mathrm{normalized}} \times \frac{P_{\mathrm{noise}}}{P(\chi=1)}.
\end{align}
This transformation enables eliminating the influence of measurement-dependent parameters, such as $\beta$ and $Q_{\rm L}$, then allowing the combination of data obtained with different parameters in the following analysis step. To search for a significant excess, the statistical uncertainty after rescaling $\sigma_{\mathrm{rescaled}}$ is defined as 
\begin{align} \sigma_{\mathrm{rescaled}} \equiv 1 \times \frac{P_{\mathrm{noise}}}{P(\chi=1)}, 
\end{align} where the factor of unity represents the standard deviation of the normalized spectrum. In the following analysis, both $P_{\rm rescaled}$ and $\sigma_{\rm rescaled}$ were subjected to the same analysis procedure.

To take into account signal width in the statistical analysis, a Maxwell-Boltzmann filter was applied by performing a cross-correlation between the processed spectrum and the Maxwell-Boltzmann distribution, which represents the expected lineshape of a dark photon signal. We assume the lineshape as
\begin{multline}
\mathcal{F}_{\text{MB}}(f) = \frac{2}{\sqrt{\pi}} \sqrt{f - f_{\text{DM}}} \left( \frac{3}{f_{\text{DM}} \frac{\braket<v^2>}{c^2} }\right)^{\frac{3}{2}}\\ \times \exp\left( -\frac{3 (f - f_{\text{DM}})}{f_{\text{DM}} \frac{\braket<v^2>}{c^2}} \right),
\end{multline}
where $f_{\rm DM}$ is the frequency of the dark photon, $c$ is the speed of light, and $\braket<v^2>$ is the mean-square velocity of the galactic halo $\braket<v^2>=(\SI{270}{km/s})^2$ \cite{PhysRevD.96.123008, cervantesADMXOrpheusFirstSearch2022}.
The systematic uncertainties, listed in Table~\ref{table:parameter} were  incorporated into the analysis; however, the statistical uncertainty was considered to be the dominant source of uncertainty \cite{cervantesADMXOrpheusFirstSearch2022}.

After these steps, all spectra were statistically combined by a weighted average to follow the maximum likelihood estimation. 
Finally, we evaluated the SNR in each frequency bin to determine whether any signal excess exceeded a 3$\sigma$ threshold (see Eq.~\ref{eq:SNR}).

\begin{figure}[htbp]
  \centering
    \includegraphics[keepaspectratio,width=0.96\linewidth]{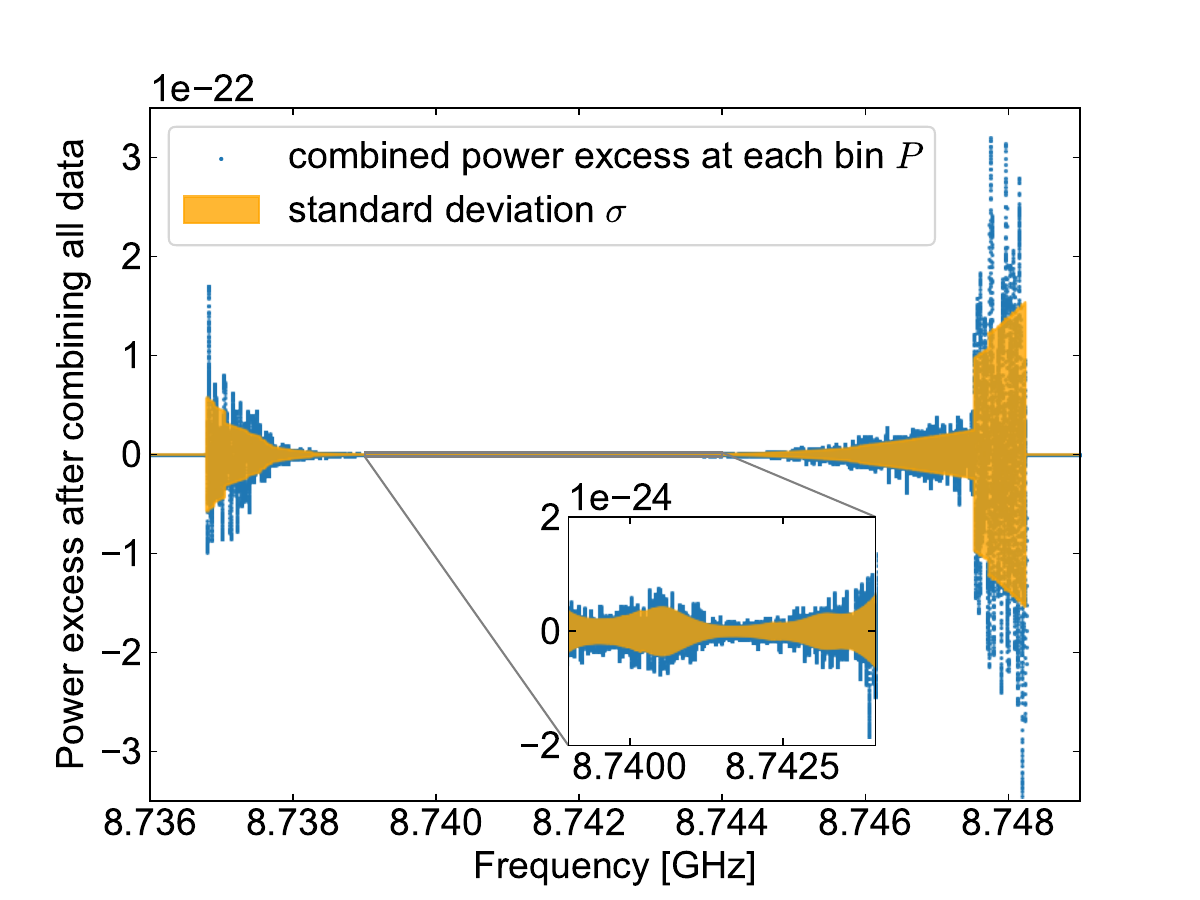}
    \caption{The combined power excesses (blue dots) and standard deviations included statistical and systematic uncertainties (orange region).}
    \label{fig:SNR}
\end{figure}

\textit{Result.---} 
Figure \ref{fig:SNR} shows the final power excess spectrum, obtained by combining all datasets, represented by blue dots. The orange region represents the standard deviation, accounting for statistical and systematic uncertainties. We found no frequency bin with a power excess exceeding $3\sigma$. Therefore, we set an upper limit on the kinetic mixing parameter. 

The unknown polarization of dark photons must also be considered in the interpretation of the experimental results. The limit on the kinetic mixing parameter, $\chi$, is consequently adjusted by a factor related to $\braket<\cos^2\theta>$, the average squared projection of the dark photon polarization vector onto the electric field of the cavity mode of interest:

\begin{align}
  \chi\rightarrow\frac{\chi}{\sqrt{\braket<\cos^2\theta>}},
\end{align}
where $\theta$ is the angle between the dark photon’s polarization vector and the sensitive axis of the experiment. Two benchmark scenarios were assumed for $\braket<\cos^2\theta>$ \cite{ariasWISPyColdDark2012,caputoDarkPhotonLimits2021}:
\begin{enumerate}
  \item Random polarization: For dark photons with no preferred polarization direction, $\braket<\cos^2\theta>=\frac{1}{3}$. 
  \item Fixed polarization for the worst case: $\braket<\cos^2\theta>=0.025$. This factor is used to set 95\% confidence level exclusion limits when the dark photon has a fixed but unknown polarization \cite{caputoDarkPhotonLimits2021}. 
\end{enumerate}

\begin{figure*}[tb] 
  \centering
    \includegraphics[keepaspectratio,width=0.8\linewidth]{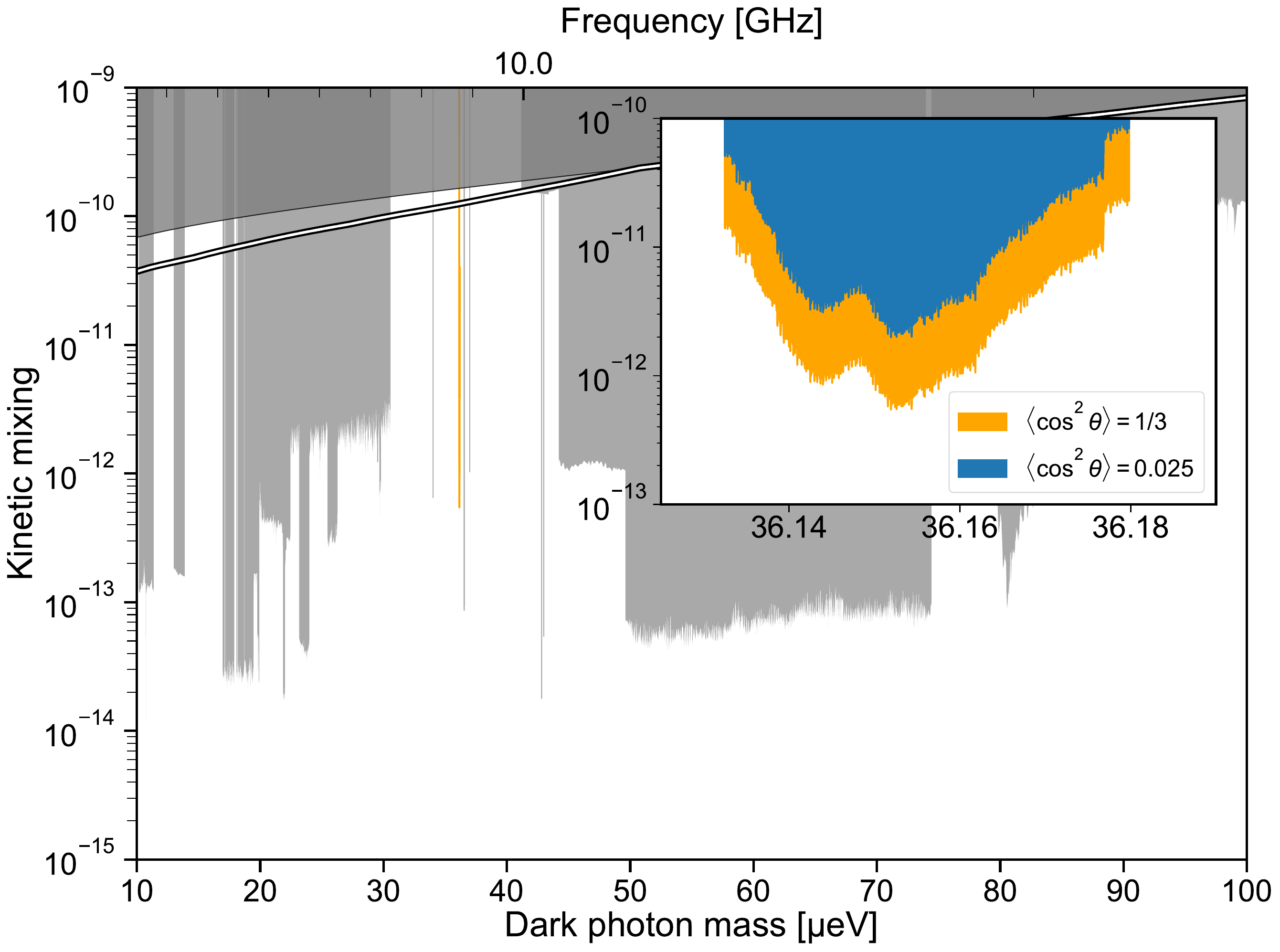}
    \caption{(Large box) Comparison with existing experimental limits and cosmological constraints, following \cite{caputoDarkPhotonLimits2021, ariasWISPyColdDark2012}. Existing results are indicated by gray shaded regions, while our result is shown as the orange line. We apply the random polarization, $\braket<\cos^2\theta>=1/3$. This plot was generated from \cite{caputoDarkPhotonLimits2021}. (Small box) The detailed view of the exclusion limit of this experiment. Orange region is in the random polarization case, and the blue region case is in the worst case \cite{caputoDarkPhotonLimits2021}.}
    \label{fig:result}
\end{figure*}

We calculated the $95\%$ confidence level upper limit of $\chi$ based on Bayesian statistics. Those two upper limits are shown in the small box of Fig.~\ref{fig:result}. 

Based on this analysis, we determined that this experiment excluded the region whose lowest limit was $\chi>5.46 \times 10^{-13}$ at a dark photon mass of $\SI{36.152}{\micro eV}$ assuming random dark photon polarization. The tuning range was $\ab[36.132, 36.179]  \ \si{\micro eV}$. Finally, we compared the result to other measurements (cosmological constraints \cite{ariasWISPyColdDark2012} and other experiments \cite{caputoDarkPhotonLimits2021}) in the \SIrange[range-units = single,range-phrase = --]{10}{100}{\micro\electronvolt} mass range (Fig.~\ref{fig:result}). 

\textit{Conclusion and prospects.---}
We have performed a cavity haloscope experiment using a tunable transmon qubit as a tuning system. This search demonstrated that the region around $\SI{36.1}{\micro eV}$ was excluded with a peak sensitivity of $\chi > 5.46\times 10^{-13}$ at \SI{36.152}{\micro eV} and the tuning range was $\ab[36.132, 36.179] \ \si{\micro eV}$, surpassing existing cosmological constraints. 

In future work, we are planning to achieve better sensitivity to the kinetic mixing parameter as well as broader frequency range of the search. To achieve higher sensitivity, improvements in the system noise temperature $T_{\rm sys}$ and the quality factor $Q_{\rm L}$ are essential. Using advanced quantum amplifiers such as JPAs or TWPAs can significantly reduce noise.
In addition, we plan to use a superconducting radio frequency cavity with a high quality factor. We expect to achieve a high $Q_{\rm L}$, up to $10^6$--$10^9$ with aluminum cavities or niobium cavities. 

There are several possibilities to expand the frequency range of the search. One approach is to utilize the strong coupling regime between the transmon qubit and the cavity. In 2D qubit-cavity systems, the coupling can be increased up to $\mathcal{O}(100) \ \si{MHz}$ \cite{PhysRevB.86.140508}. In 3D qubit-cavity systems, the previous study \cite{Noguchi_2016} demonstrated a large frequency shift with a galvanic contact between the qubit's capacitance pad and the cavity inner wall. In the next step, we are planning to utilize this technique.



\vspace{0.5 cm}
\noindent {\it Acknowledgements.---}
%
We thank H.~Fukuyama and R.~Toda for technical supports for cryogenic system and insightful discussions. 
We thank S.~Otsuka and K.~Shimozawa for metalworking and advice.
%
This work was carried out using the joint-use facilities of the Millikelvin Quantum Platform at the Cryogenic Research Center, The University of Tokyo.

%
This work was supported by “Nanotechnology Platform Japan” of the Ministry of Education, Culture, Sports, Science, and Technology (MEXT) Grant Number JPMXP1225UT1027, JPMXP1224UT1047 and fabrication was conducted in Takeda super cleanroom with the help of Nanofabrication Platform Center of School of Engineering, the University of Tokyo, Japan.

This work was supported by “Advanced Research Infrastructure for Materials and Nanotechnology in Japan (ARIM)” of the Ministry of Education, Culture, Sports, Science and Technology (MEXT) Grant Number JPMXP1223UT0001, and the fabrication was conducted in Takeda Cleanroom, Center of The University of Tokyo for The Advanced Research Infrastructure for Materials and Data Hub.

%
We are grateful for the help and support provided by Dr. ~T.~Miyazawa from the Engineering section of Core Facilities at Okinawa Institute of Science and Technology Graduate University.
%
This work made use of the Pritzker Nanofabrication Facility, part of the Pritzker School of Molecular Engineering at the University of Chicago, which receives support from Soft and Hybrid Nanotechnology Experimental (SHyNE) Resource (NSF ECCS-2025633), a node of the National Science Foundation’s National Nanotechnology Coordinated Infrastructure RRID: SCR022955.
%
This work made use of the Center of MicroNanoTechnology (CMi) at EPFL.
%
This work was supported by World Premier International Research Center Initiative (WPI), MEXT, Japan.
This work was supported by JSPS KAKENHI Grants No.~23H01182, 23K13093, 23H04864, 23KJ0678, 23K17688, 23K22486, 24K17042, 24H00689, 25H00638, 25K01610, 25KJ0847, JST ASPIRE Grant No. JPMJAP2316, JST PRESTO Grant No.~JPMJPR2253, JPMJPR23F7, JST SPRING JPMJSP2108, Japan.
%
QUP Release No.~KEK-QUP-2025-0011.

\bibliography{references}





\end{document}